\documentclass[%
 reprint,
superscriptaddress,
 amsmath,amssymb,
 aps,
]{revtex4-2}

\usepackage{graphicx}% Include figure files
\usepackage{dcolumn}% Align table columns on decimal point
\usepackage{bm}% bold math
\begin{document}

\preprint{APS/123-QED}

\title{Spatial resolution of omni-resonant imaging}
% Force line breaks with \\

\author{Abbas Shiri}
\affiliation{CREOL, The College of Optics \& Photonics, University of Central Florida, Orlando, FL 32816, USA}
\affiliation{Department of Electrical and Computer Engineering, University of Central Florida, Orlando, FL 32816, USA}
\author{Ayman F. Abouraddy}
\affiliation{CREOL, The College of Optics \& Photonics, University of Central Florida, Orlando, FL 32816, USA}
\affiliation{Department of Electrical and Computer Engineering, University of Central Florida, Orlando, FL 32816, USA}
\email{*raddy@creol.ucf.edu}

%\affiliation{%


%\affiliation{
%\affiliation{
%\affiliation{%


%\noindent\textbf{Keywords.} Space-time supermode, multi-mode waveguide optics, group-velocity, non-diffraction

\begin{abstract}
Omni-resonance refers to the broadening of the spectral transmission through a planar cavity, not by changing the cavity structure, but by judiciously preconditioning the incident optical field. As such, broadband imaging can be performed through such a cavity with all the wavelengths simultaneously resonating. We examine here the spatial resolution of omni-resonant imaging and find that the spectral linewidth of the cavity resonance determines the spatial resolution. Surprisingly, the spatial resolution improves at longer wavelengths because of the negative angular dispersion intrinsic to Fabry-P{\'e}rot resonances, in contrast to conventional diffraction-limited optical imaging systems where the spatial resolution improves at shorter wavelengths. These results are important for applications ranging from transparent solar windows to nonlinear resonant image processing.
\end{abstract}

%\setboolean{displaycopyright}{true}

\maketitle

Light resonates with a planar Fabry-P{\'e}rot (FP) cavity only in the vicinity of discrete wavelengths. A host of benefits ensue from resonant field buildup in a cavity; e.g., enhancement of linear \cite{Chong10PRL,Wan11S,Villinger15OL,Baranov17NRM} and nonlinear \cite{Makri14PRA} absorption. However, such resonant enhancements are narrowband phenomena harnessed only at the discrete resonances. To deliver such enhancements over broad bandwidths, efforts have been dedicated to increasing the resonant linewidth \textit{without} reducing the cavity finesse. One example is so-called `white-light' cavities \cite{Wicht97OC,Rinkleff05PS} that require introducing a medium (e.g., atomic \cite{Pati07PRL,Wu08PRA} or nonlinear \cite{Yum13JLT} media) whose dispersion characteristics change the resonant response. To date, introducing linear components into the cavity (e.g., grating pairs \cite{Wise05PRL} or chirped Bragg mirrors \cite{Yum13OC}) has not yielded the desired effect. Another approach makes use of trapdoor strategies \cite{Xu06NPhys}, but these are complex and typically require further optical sources for their operation.

We have recently investigated an `omni-resonant' configuration \cite{Shabahang17SR} in which broadband light resonates continuously with a planar Fabry-P{\'e}rot (FP) cavity without modifying the cavity in any way. Instead, by pre-conditioning an incident collimated field and introducing angular dispersion matching that of one of the cavity resonances, this resonance becomes `achromatic', and its bandwidth not only can exceed the resonant linewidth, but can even exceed the cavity free spectral range (FSR) \cite{Shabahang17SR}. All wavelengths within a broad continuous spectrum couple to the cavity and resonate with it, thus effectively decoupling the resonance linewidth from the cavity-photon lifetime. Resonantly enhanced linear and nonlinear optical effects can be consequently harnessed over a broad bandwidth. Indeed, we have utilized omni-resonance to realize broadband resonant coherent perfect absorption \cite{Villinger21AOM,Jahromi21arxiv}. Omni-resonance is not to be confused with omni-directional resonances in metal-dielectric-metal structures in which a wavelength resonates with the structure regardless of its incident angle \cite{Shin04APL,Liu07APL}.

A broad range of potential applications motivate extending the concept of omni-resonance to optical imaging configurations. First, omni-resonance can be exploited in transparent solar windows in which only a particular spectral band is absorbed. Indeed, we have recently demonstrated a doubling in the near-infrared external quantum efficiency in an ultra-thin amorphous silicon PIN diode in a planar cavity \cite{Villinger21AOM}. Second, an omni-resonant cavity can toggle between broadband \textit{passive} imaging with incoherent light and \textit{active} imaging with a narrowband laser \cite{Shabahang19OL}. Furthermore, resonantly nonlinear effects with an intense laser field \cite{Makri14PRA} can be combined with broadband low-intensity imaging. For example, ultrashort `space-time' wave packets \cite{Kondakci17NP,Kondakci19NC} -- a recently developed class of propagation-invariant pulsed beams \cite{Yessenov22AOP} -- were realized in the omni-resonant condition, so that their entire bandwidth \cite{Shiri20OL} or selected segments of it \cite{Shiri20APLP} can resonate with the cavity. It is thus crucial to determine the spatial resolution of omni-resonant imaging in order to evaluate the limits of these applications.

In this paper, we present a theoretical formulation of the spatial-resolution limit in omni-resonant imaging and provide experimental validation of our predictions. Because of the space-time coupling associated with omni-resonance, we find that the \textit{resonant spectral linewidth} of a planar FP cavity determines the \textit{spatial resolution} of omni-resonant imaging through it. Using a standard Air Force resolution chart, we evaluate the spatial resolution of omni-resonant imaging in FP cavities of different finesse and obtain quantitative validation of our model. 

\begin{figure}[t!]
\centering
\includegraphics[width=8.1cm]{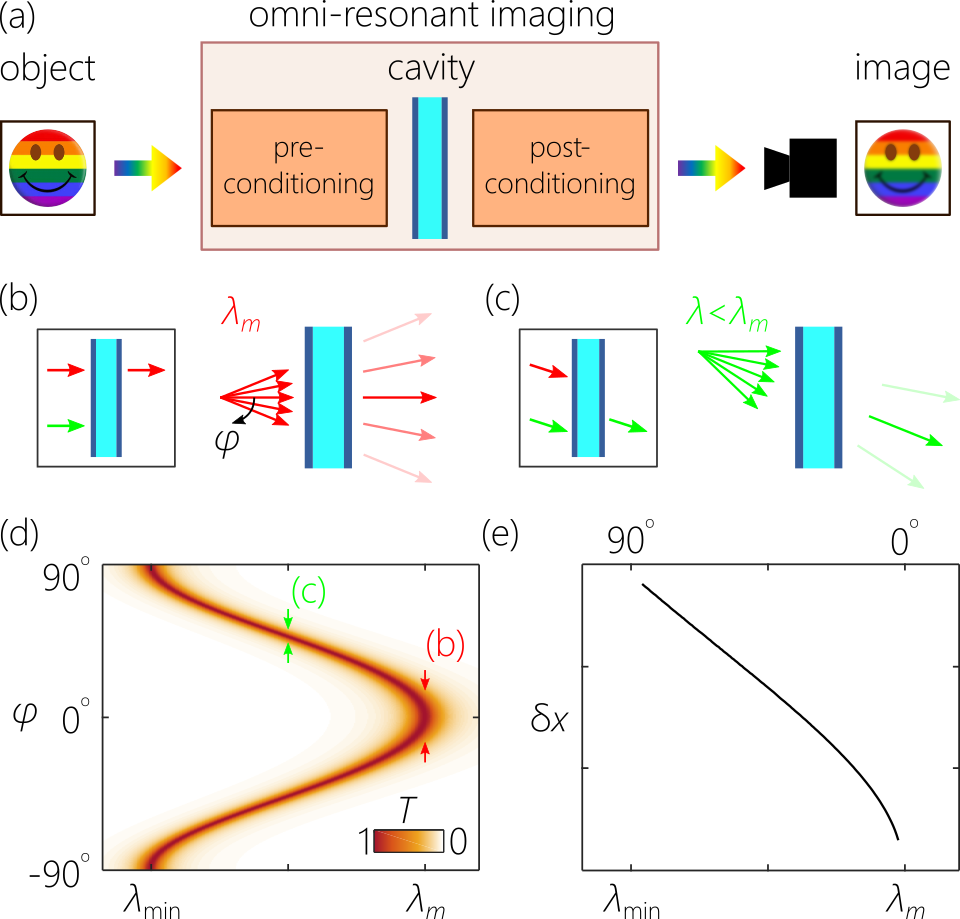}
\caption{(a) Concept of omni-resonant imaging through a planar FP cavity. (b) At normal incidence, the wavelength $\lambda_{m}$ associated with the $m^{\mathrm{th}}$-order resonance is transmitted. At $\lambda_{m}$, a narrow angular bandwidth centered on $\varphi\!=\!0^{\circ}$ is transmitted. (c) At oblique incidence, a blue-shifted wavelength $\lambda\!<\!\lambda_{m}$ resonates with the cavity, and a narrow angular bandwidth centered on $\varphi\!>\!0^{\circ}$ is transmitted. (d) Plot of the intensity transmission $T(\lambda,\varphi)$ for the $m^{\mathrm{th}}$-order resonance for a planar FP cavity of finite finesse. Points along the angular dispersion curve corresponding to the configurations in (b) and (c) are indicated. (e) Plot of the spatial resolution $\delta x(\lambda)$ (Eq.~\ref{Eq:SpatialResolution}) obtained from the angular dispersion curve in (d). The top axis is the incident angle $\varphi(\lambda)$ at the exact resonance condition.}
\label{Fig:concept}
\end{figure}

The general configuration we call `omni-resonant imaging' is depicted schematically in Fig.~\ref{Fig:concept}(a). The object is illuminated with broadband light, and the scattered field is pre-conditioned before it traverses a planar FP cavity. The resonantly transmitted light is then post-conditioned (to invert the pre-conditioning), and an image is formed \cite{Shabahang17SR,Shabahang19OL}. In this omni-resonant imaging system [Fig.~\ref{Fig:concept}(a)], the entire spectrum is transmitted through the cavity, and a broadband image is formed, in principle with the same bandwidth as the incident radiation. We pose here the following question: what is the spatial-resolution limit of the broadband image formed after imposing the omni-resonant condition? 

To formulate the answer to this question, we consider the spectral transmission at a free-space wavelength $\lambda$ and external incident angle $\varphi$ from free space with respect to the normal to the cavity \cite{SalehBook07}: $T(\lambda,\varphi)\!=\!1/\{1+(\tfrac{2F}{\pi})^{2}\sin^{2}{\Psi}\}$, where $F$ is the cavity finesse, $\psi$ is the round-trip phase, $\psi\!=\!\tfrac{4\pi d}{\lambda}\sqrt{n^{2}-\sin^{2}{\phi}}$, $d$ is the thickness of the cavity defect layer, $n$ its refractive index, and we have neglected the reflection phases from the two cavity mirrors. The resonant wavelength at normal incidence $\varphi\!=\!0$ for the $m^{\mathrm{th}}$-order resonance is $\lambda_{m}\!=\!\tfrac{2nd}{m}$ [Fig.~\ref{Fig:concept}(b)], which blue-shifts ($\lambda\!<\!\lambda_{m}$) when $\varphi\!>\!0$ [Fig.~\ref{Fig:concept}(c)], reaching $\lambda_{\mathrm{min}}\!=\!\tfrac{2d}{m}\sqrt{n^{2}-1}$ when $\varphi\!\rightarrow\!90^{\circ}$. Plotting the FP cavity transmission $T(\lambda,\varphi)$ in Fig.~\ref{Fig:concept}(d) reveals that the resonance is associated with angular dispersion: at each incident angle $\varphi$, a particular wavelength $\lambda$ is transmitted, $\tfrac{\lambda}{\lambda_{m}}\!=\!\sqrt{1-\tfrac{1}{n^{2}}\sin^{2}{\varphi}}$. The finite cavity finesse relaxes the exact resonant condition, thereby permitting a narrow spectral linewidth at each incident angle.

Note that the angular dispersion associated with any FP resonance is negative; that is, the required incident angle drops with wavelength $\tfrac{d\varphi}{d\lambda}\!<\!0$. Although gratings provide positive $\tfrac{d\varphi}{d\lambda}$ for normally incident light, changing the geometry can effectively deliver negative $\tfrac{d\varphi}{d\lambda}$ \cite{England14PNAS,Lin15OL,Shabahang17SR}. More recently, it has been shown that a meta-surface or cascade of meta-surfaces can control the sign of $\tfrac{d\varphi}{d\lambda}$ \cite{Arbabi17Optica,McClung20Light}. In all cases, however, the magnitude of $\tfrac{d\varphi}{d\lambda}$ produced does not exceed that yielded by a conventional grating -- an obstacle that we tackle below.    

As mentioned above, the cavity finesse produces a narrow spectral linewidth centered upon the ideal resonant wavelength at any given incident angle. Alternatively, at each wavelength $\lambda$ (between $\lambda_{\mathrm{min}}$ and $\lambda_{m}$) a narrow angular bandwidth $\delta\varphi$ centered at an external incident angle $\varphi$ can resonantly transmit through the cavity, where $\sin{\varphi}\!=\!\pm n\sqrt{1-(\tfrac{\lambda}{\lambda_{m}})^{2}}$. The angular bandwidth $\delta\varphi(\lambda)$ centered on $\varphi(\lambda)$ is $\delta\varphi(\lambda)\!=\!\varphi_{+}-\varphi_{-}$, where
\begin{equation}
\sin{\varphi_{\pm}}=n\sqrt{1-\left\{(1\mp\eta)\left(\frac{\lambda}{\lambda_{m}}\right)\right\}^{2}};
\end{equation}
here $\eta\!=\!\tfrac{1}{m\pi}\sin^{-1}(\tfrac{\pi}{2F})$ is a dimensionless parameter. For a FP cavity of a particular thickness $d$ and index $n$, and for the $m^{\mathrm{th}}$-order resonance, the spatial resolution $\delta x(\lambda)$ is estimated as:
\begin{equation}\label{Eq:SpatialResolution}
\delta x(\lambda)=\frac{\lambda}{\sin{\varphi_{+}}-\sin{\varphi_{-}}};
\end{equation}
which we plot in Fig.~\ref{Fig:concept}(e). The spatial resolution $\delta x(\lambda)$ is the size of the smallest discernible feature along $x$ at wavelength $\lambda$, which is associated with a particular incident angle $\varphi(\lambda)$. Intriguingly, $\delta x(\lambda)$ decreases with $\lambda$ (the resolution improves at longer wavelengths), in contrast to conventional diffraction-limited optical imaging systems in which $\delta x(\lambda)$ increases with $\lambda$ (the resolution deteriorates at longer wavelengths). This behavior is a consequence of the negative angular dispersion associated with planar FP cavity resonances [Fig.~\ref{Fig:concept}(d)].

\begin{figure}[t!]
\centering
\includegraphics[width=8.1cm]{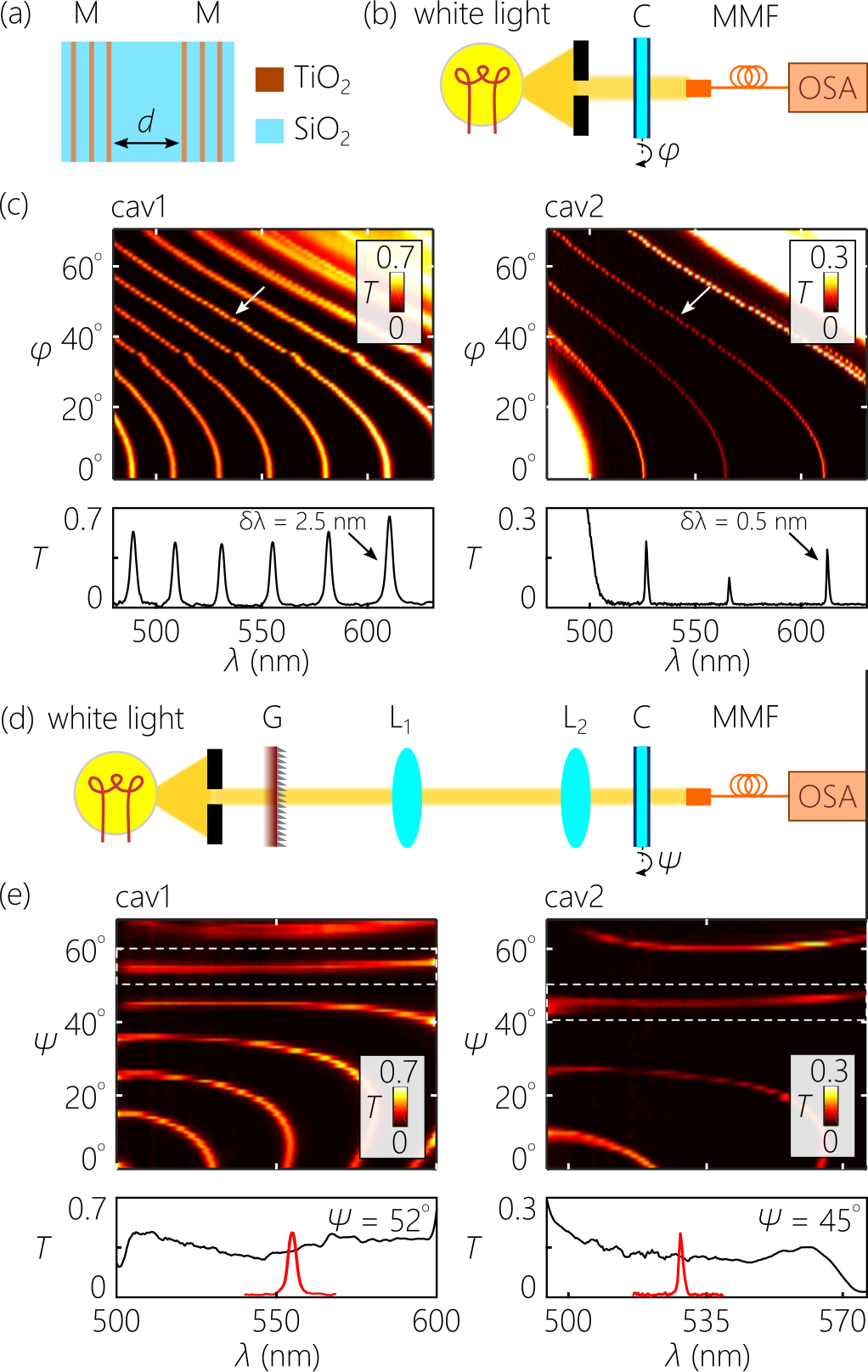}
\caption{(a) Structure of the FP cavities used (M: Bragg mirror). (b) Setup for measuring $T(\lambda,\varphi)$; all the wavelengths in the spectrum are incident on the cavity at the same angle. C: FP cavity; MMF: multimode optical fiber; OSA: optical spectrum analyzer. (c) Measured $T(\lambda,\varphi)$ for cav1 and cav2. Only discrete resonant wavelengths are transmitted at any $\varphi$. Bottom panels are the normal incident transmission spectra $T(\lambda,0^{\circ})$, and $\delta\lambda$ is the resonant linewidth. (d) Setup for realizing the omni-resonant condition. Each wavelength is directed at a different angle, and the cavity is rotated an angle $\psi$ around its axis. G: Diffraction grating; L: lens. (e) Measured $T(\lambda,\psi)$ for cav1 and cav2 in the omni-resonant configuration. The bottom panels are the achromatic resonances emerging at $\psi\!\approx\!52^{\circ}$ for cav1 (corresponding to the dashed box in the upper panel), and at $\psi\!\approx\!45^{\circ}$ for cav2. Normal-incidence resonances in the same spectral range are plotted for comparison.}
\label{Fig:OmniResonance}
\end{figure}

We make use of two optical cavities in our experiments, each is deposited on a glass substrate and consists of two symmetric Bragg mirrors sandwiching a silica defect layer of thickness $d$ [Fig.~\ref{Fig:OmniResonance}(a)]. The Bragg mirrors (deposited via ebeam evaporation) comprise bilayers of TiO$_2$ ($n\!=\!2.28$) and SiO$_2$ ($n\!=\!1.49$). For the first cavity (referred to hereon as `cav1'), each Bragg mirror is formed of 5 bilayers and has a reflectivity of $R\!=\!0.82$, corresponding to a finesse of $\sim\!18$, and the defect-layer thickness is $d\!\approx\!4.15$~$\mu$m. The Bragg mirror for the second cavity (`cav2') is formed of 8 bilayers, $R\!=\!0.98$, a finesse of $\sim\!90$, and $d\!=\!2.2$~$\mu$m. The measured resonance linewidth for cav1 is 2.5~nm and for cav2 is 0.5~nm, both at a wavelength $\sim\!610$~nm.

We first acquire the angular-dispersion curve for the cavity resonant transmission $T(\lambda,\varphi)$ using a collimated white-light source (Thorlabs QTH10). The transmitted light is collected by a multi-mode fiber (300-$\mu$m core diameter) and delivered to a spectrometer (Jaz, Ocean Optics); see Fig.~\ref{Fig:OmniResonance}(b). The measurement results are plotted in Fig.~\ref{Fig:OmniResonance}(c) for cav1 and cav2 while rotating each cavity around its axis an angle $\varphi$ in $1^{\circ}$-steps. The slope in the linear portion of the angular-dispersion curve for the resonances indicated in Fig.~\ref{Fig:OmniResonance}(c) -- identified by a small white arrow -- is $-0.39^{\circ}$/nm for cav1, and $-0.41^{\circ}$/nm for cav2. These angular-dispersion values are critical for designing the omni-resonant imaging setup described next.

For omni-resonant imaging, we introduce into the incident field the same angular dispersion associated with the selected FP resonance. This process of angular-dispersion matching yields an `achromatic resonance', whereby an extended incident spectrum resonates continuously with the cavity and is transmitted. To realize this omni-resonant configuration experimentally, the spectrum of the collimated white-light source used in Fig.~\ref{Fig:OmniResonance}(a) is angularly resolved with a transmission grating with a line density of 1400~lines/mm (at an incident angle of $17^{\circ}$ and $8^{\circ}$ for cav1 and cav2, respectively) before incidence on the FP cavity [Fig.~\ref{Fig:OmniResonance}(d)]. The angular dispersion produced by the grating is almost linear with values $\tfrac{d\varphi}{d\lambda}\!\approx\!-0.097$ and $-0.102^{\circ}$/nm, which are $\approx\!4\times$ smaller than that of the FP resonances in Fig.~\ref{Fig:OmniResonance}(c). An imaging system comprising lenses L$_1$ and $L_2$ of focal lengths 100~mm and 25~mm [Fig.~\ref{Fig:OmniResonance}(d)], respectively, provides a $4\times$ enhancement in the magnitude of the angular dispersion as required for the omni-resonant condition. Now each wavelength is incident on the cavity at a slightly different angle. Taking the optical axis to correspond to $\lambda\!=\!540$~nm and rotating the cavity an angle $\psi$ around this axis, we obtain the angular dispersion curves in Fig.~\ref{Fig:OmniResonance}(e) for cav1 and cav2. At particular values of the rotation angle $\psi$, the spectral transmission is flat; that is, a broadband continuous spectrum is transmitted rather than discrete resonances. For cav1, the initial resonant transmission bandwidth of $\sim\!2.5$~nm is extended to $\sim\!100$~nm, and for cav2 from $\sim\!0.5$~nm to $\sim\!70$~nm (the spectra here are limited by the acceptance angles of the lenses L$_1$ and L$_2$).

\begin{figure}[t!]
\centering
\includegraphics[width=8.6cm]{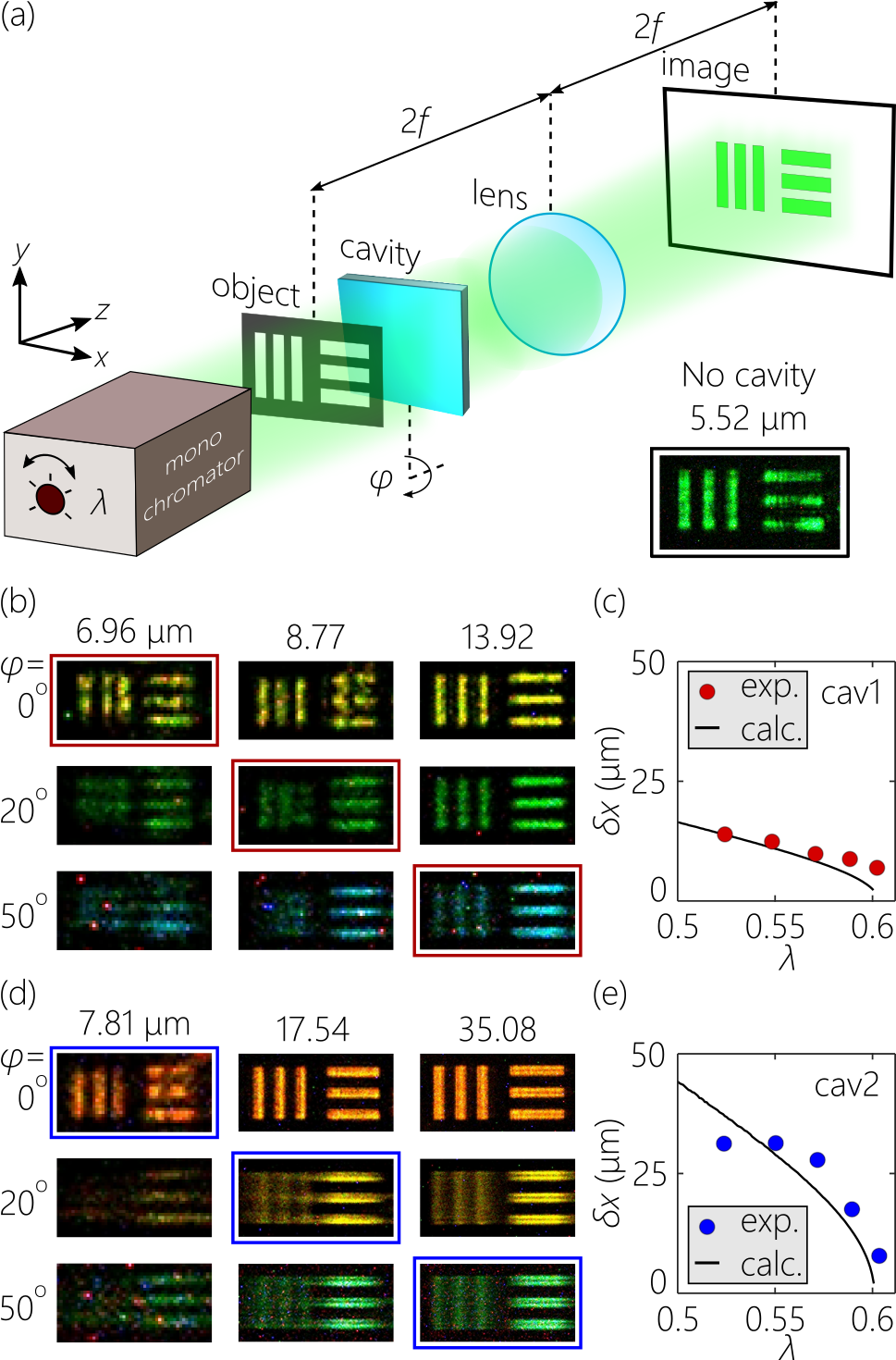}
\caption{(a) Schematic of the setup for determining the spatial resolution of omni-resonant imaging. The inset at the bottom right helps identify the spatial resolution of the imaging system in absence of the cavity. (b) Samples of the recorded images in presence of cav1 for chart lines of different separation (identified on the top) at three angular settings for $\phi$. The panels with highlighted borders identify the spatial resolution limit at this angle $\varphi(\lambda)$. (c) Measured (dots) wavelength-dependent spatial resolution $\delta x(\lambda)$ compared the theoretical prediction (solid curve). (d,e) Same as (b,c) for cav2.}
\label{Fig:Spatial resolution}
\end{figure}

Making use of these measurements, we assess the wavelength-dependent spatial resolution $\delta x(\lambda)$ of omni-resonant imaging in cav1 and cav2 using as the object the 1951 USAF resolution test chart [Fig.~\ref{Fig:Spatial resolution}(a)]. Using a home-built monochromator that delivers a wavelength-tunable plane wave with $\approx\!1$-nm spectral resolution across the visible spectrum, we illuminate the object, and the transmitted light impinges on the cavity, which is rotated an angle $\varphi(\lambda)$ at each wavelength $\lambda$ as determined by the omni-resonant condition as depicted in Fig.~\ref{Fig:OmniResonance}(c,e). An achromatic lens (focal length $f\!=\!40$~mm) images the object plane to a CCD camera (ImagingSource, DFK 33UX178) with unity magnification ($d_{1}\!=\!d_{2}\!=\!80$~mm), and the cavity is placed midway between the object and the lens (the results are not sensitive to the positioning of the cavity between object and lens). At normal incidence for both cavities, the resonant wavelength is $\approx\!611$~nm, and rotating the cavity from $0^{\circ}$ to $\approx\!60^{\circ}$ spans the spectrum down to $\approx\!500$~nm.

At each wavelength/angle setting of the source and cavity, we capture an image of the object and determine the spatial resolution $\delta x(\lambda)$. We estimate $\delta x(\lambda)$ to be the separation between the lines when the value of the intensity recorded in the valley between them reaches half that of the peak intensity of the lines themselves. We consider that the lines are not resolved beyond this condition. Examples are shown for cav1 in Fig.~\ref{Fig:Spatial resolution}(b), where we plot the portion of the test chart corresponding to the resolution criterion at different cavity tilt angles (and thus different incident wavelengths). For example, at $\varphi\!=\!0^{\circ}$ for cav1 [Fig.~\ref{Fig:Spatial resolution}(b)], features separated by 6.96~$\mu$m are at the spatial-resolution threshold, whereas larger features are resolved. At $\varphi\!=\!20^{\circ}$, on the other hand, 8.77-$\mu$m features are at the spatial-resolution limit; larger features (13.92~$\mu$m) are resolved whereas smaller features (6.96~$\mu$m) are \textit{not}.

Following this approach, we obtain the spatial resolution $\delta x(\lambda)$ for cav1 and plot the results in Fig.~\ref{Fig:Spatial resolution}(c). The corresponding results for cav2 are plotted in Fig.~\ref{Fig:Spatial resolution}(d,e). We note that $\delta x(\lambda)$ drops at longer wavelengths as predicted, in contrast to conventional diffraction-limited imaging systems where the spatial resolution worsens at longer wavelengths. Furthermore, it is clear that $\delta x(\lambda)$ for the lower-finesse cav1 is superior to that of the higher-finesse cav2. The theoretical curves are obtained from Eq.~\ref{Eq:SpatialResolution} after introducing an overall multiplicative factor as a fitting parameter. The measured $\delta x(\lambda)$ is smaller than the theoretical predictions, which we attribute to the fact that the cavity finesses are likely lower than expected theoretically based on the Bragg mirror structure (because of limitations of manufacturing precision) and to the finite spectral uncertainty of the illumination source (the monochromator).  

In our work here, we have assumed that the spatial resolution limit of omni-resonant imaging is set by the cavity, and that $\delta x(\lambda)$ overrides the resolution limit of the imaging system with the cavity removed. This assumption is verified in Fig.~\ref{Fig:Spatial resolution}(b,d) where the images of the vertical and horizontal lines are quite distinct. Because the grating spreads the wavelength along one axis, and we rotate the cavity around the axis parallel to the grating, the omni-resonant imaging limit applies only along that direction, corresponding to the vertical lines in Fig.~\ref{Fig:Spatial resolution}(b,d). The spatial resolution along the direction of the horizontal lines in Fig.~\ref{Fig:Spatial resolution}(b,d) is determined by the optical system, and is estimated at $\delta x\!\approx\!2.5$~$\mu$ at $\lambda\!=\!500$~nm [see also Fig.~\ref{Fig:Spatial resolution}(a), inset]. 

In conclusion, we have estimated the spatial resolution of omni-resonant imaging and verified the predictions experimentally using two planar FP cavities of different finesse. Omni-resonance refers to the condition whereupon a broadband optical field is pre-conditioned by introducing the angular dispersion, after which the field resonates with the cavity continuously across its entire spectrum. Extended to the context of imaging, omni-resonance allows for broadband light to be used to image an object through a cavity without spectral filtering. The spatial resolution of omni-resonant imaging is dictated by the cavity finesse; the smallest discernible feature size is proportional to the finesse (high-finesse cavities have large $\delta x$, and vice versa). Therefore, increasing the resonant field buildup in the cavity to enhance nonlinear interactions will be associated with a loss of spatial resolution. These results will be critical for optical signal processing schemes that combine linear and nonlinear interactions in a cavity. 

%\begin{backmatter}
\section*{\textbf{Funding}}
U.S. Office of Naval Research (ONR) N00014-17-1-2458 and N00014-20-1-2789.

\section*{\textbf{Disclosures}}
The authors declare no conflicts of interest.

%\bigskip
%\end{backmatter}

\bibliography{diffraction}

\begin{thebibliography}{29}%
\makeatletter
\providecommand \@ifxundefined [1]{%
 \@ifx{#1\undefined}
}%
\providecommand \@ifnum [1]{%
 \ifnum #1\expandafter \@firstoftwo
 \else \expandafter \@secondoftwo
 \fi
}%
\providecommand \@ifx [1]{%
 \ifx #1\expandafter \@firstoftwo
 \else \expandafter \@secondoftwo
 \fi
}%
\providecommand \natexlab [1]{#1}%
\providecommand \enquote  [1]{``#1''}%
\providecommand \bibnamefont  [1]{#1}%
\providecommand \bibfnamefont [1]{#1}%
\providecommand \citenamefont [1]{#1}%
\providecommand \href@noop [0]{\@secondoftwo}%
\providecommand \href [0]{\begingroup \@sanitize@url \@href}%
\providecommand \@href[1]{\@@startlink{#1}\@@href}%
\providecommand \@@href[1]{\endgroup#1\@@endlink}%
\providecommand \@sanitize@url [0]{\catcode `\\12\catcode `\$12\catcode
  `\&12\catcode `\#12\catcode `\^12\catcode `\_12\catcode `\%12\relax}%
\providecommand \@@startlink[1]{}%
\providecommand \@@endlink[0]{}%
\providecommand \url  [0]{\begingroup\@sanitize@url \@url }%
\providecommand \@url [1]{\endgroup\@href {#1}{\urlprefix }}%
\providecommand \urlprefix  [0]{URL }%
\providecommand \Eprint [0]{\href }%
\providecommand \doibase [0]{https://doi.org/}%
\providecommand \selectlanguage [0]{\@gobble}%
\providecommand \bibinfo  [0]{\@secondoftwo}%
\providecommand \bibfield  [0]{\@secondoftwo}%
\providecommand \translation [1]{[#1]}%
\providecommand \BibitemOpen [0]{}%
\providecommand \bibitemStop [0]{}%
\providecommand \bibitemNoStop [0]{.\EOS\space}%
\providecommand \EOS [0]{\spacefactor3000\relax}%
\providecommand \BibitemShut  [1]{\csname bibitem#1\endcsname}%
\let\auto@bib@innerbib\@empty
%</preamble>
\bibitem [{\citenamefont {Chong}\ \emph {et~al.}(2010)\citenamefont {Chong},
  \citenamefont {Ge}, \citenamefont {Cao},\ and\ \citenamefont
  {Stone}}]{Chong10PRL}%
  \BibitemOpen
  \bibfield  {author} {\bibinfo {author} {\bibfnamefont {Y.~D.}\ \bibnamefont
  {Chong}}, \bibinfo {author} {\bibfnamefont {L.}~\bibnamefont {Ge}}, \bibinfo
  {author} {\bibfnamefont {H.}~\bibnamefont {Cao}},\ and\ \bibinfo {author}
  {\bibfnamefont {A.~D.}\ \bibnamefont {Stone}},\ }\bibfield  {title} {\bibinfo
  {title} {Coherent perfect absorbers: Time-reversed lasers},\ }\href@noop {}
  {\bibfield  {journal} {\bibinfo  {journal} {Phys. Rev. Lett.}\ }\textbf
  {\bibinfo {volume} {105}},\ \bibinfo {pages} {053901} (\bibinfo {year}
  {2010})}\BibitemShut {NoStop}%
\bibitem [{\citenamefont {Wan}\ \emph {et~al.}(2011)\citenamefont {Wan},
  \citenamefont {Chong}, \citenamefont {Ge}, \citenamefont {Noh}, \citenamefont
  {Stone},\ and\ \citenamefont {Cao}}]{Wan11S}%
  \BibitemOpen
  \bibfield  {author} {\bibinfo {author} {\bibfnamefont {W.}~\bibnamefont
  {Wan}}, \bibinfo {author} {\bibfnamefont {Y.}~\bibnamefont {Chong}}, \bibinfo
  {author} {\bibfnamefont {L.}~\bibnamefont {Ge}}, \bibinfo {author}
  {\bibfnamefont {H.}~\bibnamefont {Noh}}, \bibinfo {author} {\bibfnamefont
  {A.~D.}\ \bibnamefont {Stone}},\ and\ \bibinfo {author} {\bibfnamefont
  {H.}~\bibnamefont {Cao}},\ }\bibfield  {title} {\bibinfo {title}
  {Time-reversed lasing and interferometric control of absorption},\
  }\href@noop {} {\bibfield  {journal} {\bibinfo  {journal} {Science}\ }\textbf
  {\bibinfo {volume} {331}},\ \bibinfo {pages} {889} (\bibinfo {year}
  {2011})}\BibitemShut {NoStop}%
\bibitem [{\citenamefont {Villinger}\ \emph {et~al.}(2015)\citenamefont
  {Villinger}, \citenamefont {Bayat}, \citenamefont {Pye},\ and\ \citenamefont
  {Abouraddy}}]{Villinger15OL}%
  \BibitemOpen
  \bibfield  {author} {\bibinfo {author} {\bibfnamefont {M.~L.}\ \bibnamefont
  {Villinger}}, \bibinfo {author} {\bibfnamefont {M.}~\bibnamefont {Bayat}},
  \bibinfo {author} {\bibfnamefont {L.~N.}\ \bibnamefont {Pye}},\ and\ \bibinfo
  {author} {\bibfnamefont {A.~F.}\ \bibnamefont {Abouraddy}},\ }\bibfield
  {title} {\bibinfo {title} {Analytical model for coherent perfect absorption
  in one-dimensional photonic structures},\ }\href@noop {} {\bibfield
  {journal} {\bibinfo  {journal} {Opt. Lett.}\ }\textbf {\bibinfo {volume}
  {40}},\ \bibinfo {pages} {5550} (\bibinfo {year} {2015})}\BibitemShut
  {NoStop}%
\bibitem [{\citenamefont {Baranov}\ \emph {et~al.}(2017)\citenamefont
  {Baranov}, \citenamefont {Krasnok}, \citenamefont {Shegai}, \citenamefont
  {Al{\'u}},\ and\ \citenamefont {Chong}}]{Baranov17NRM}%
  \BibitemOpen
  \bibfield  {author} {\bibinfo {author} {\bibfnamefont {D.~G.}\ \bibnamefont
  {Baranov}}, \bibinfo {author} {\bibfnamefont {A.}~\bibnamefont {Krasnok}},
  \bibinfo {author} {\bibfnamefont {T.}~\bibnamefont {Shegai}}, \bibinfo
  {author} {\bibfnamefont {A.}~\bibnamefont {Al{\'u}}},\ and\ \bibinfo {author}
  {\bibfnamefont {Y.}~\bibnamefont {Chong}},\ }\bibfield  {title} {\bibinfo
  {title} {Coherent perfect absorbers: linear control of light with light},\
  }\href@noop {} {\bibfield  {journal} {\bibinfo  {journal} {Nat. Rev. Mater.}\
  }\textbf {\bibinfo {volume} {2}},\ \bibinfo {pages} {17064} (\bibinfo {year}
  {2017})}\BibitemShut {NoStop}%
\bibitem [{\citenamefont {Makri}\ \emph {et~al.}(2014)\citenamefont {Makri},
  \citenamefont {Ramezani}, \citenamefont {Kottos},\ and\ \citenamefont
  {Vitebskiy}}]{Makri14PRA}%
  \BibitemOpen
  \bibfield  {author} {\bibinfo {author} {\bibfnamefont {E.}~\bibnamefont
  {Makri}}, \bibinfo {author} {\bibfnamefont {H.}~\bibnamefont {Ramezani}},
  \bibinfo {author} {\bibfnamefont {T.}~\bibnamefont {Kottos}},\ and\ \bibinfo
  {author} {\bibfnamefont {I.}~\bibnamefont {Vitebskiy}},\ }\bibfield  {title}
  {\bibinfo {title} {Concept of a reflective power limiter based on nonlinear
  localized modes},\ }\href@noop {} {\bibfield  {journal} {\bibinfo  {journal}
  {Phys. Rev. A}\ }\textbf {\bibinfo {volume} {89}},\ \bibinfo {pages}
  {031802(R)} (\bibinfo {year} {2014})}\BibitemShut {NoStop}%
\bibitem [{\citenamefont {Wicht}\ \emph {et~al.}(1997)\citenamefont {Wicht},
  \citenamefont {Danzmann}, \citenamefont {Fleischhauer}, \citenamefont
  {Scully}, \citenamefont {M{\"u}ller},\ and\ \citenamefont
  {Rinkleff}}]{Wicht97OC}%
  \BibitemOpen
  \bibfield  {author} {\bibinfo {author} {\bibfnamefont {A.}~\bibnamefont
  {Wicht}}, \bibinfo {author} {\bibfnamefont {K.}~\bibnamefont {Danzmann}},
  \bibinfo {author} {\bibfnamefont {M.}~\bibnamefont {Fleischhauer}}, \bibinfo
  {author} {\bibfnamefont {M.}~\bibnamefont {Scully}}, \bibinfo {author}
  {\bibfnamefont {G.}~\bibnamefont {M{\"u}ller}},\ and\ \bibinfo {author}
  {\bibfnamefont {R.-H.}\ \bibnamefont {Rinkleff}},\ }\bibfield  {title}
  {\bibinfo {title} {White-light cavities, atomic phase coherence, and
  gravitational wave detectors},\ }\href@noop {} {\bibfield  {journal}
  {\bibinfo  {journal} {Opt. Commun.}\ }\textbf {\bibinfo {volume} {134}},\
  \bibinfo {pages} {431} (\bibinfo {year} {1997})}\BibitemShut {NoStop}%
\bibitem [{\citenamefont {Rinkleff}\ and\ \citenamefont
  {Wicht}(2005)}]{Rinkleff05PS}%
  \BibitemOpen
  \bibfield  {author} {\bibinfo {author} {\bibfnamefont {R.-H.}\ \bibnamefont
  {Rinkleff}}\ and\ \bibinfo {author} {\bibfnamefont {A.}~\bibnamefont
  {Wicht}},\ }\bibfield  {title} {\bibinfo {title} {The concept of white light
  cavities using atomic phase coherence},\ }\href@noop {} {\bibfield  {journal}
  {\bibinfo  {journal} {Phys. Scr.}\ }\textbf {\bibinfo {volume} {2005}},\
  \bibinfo {pages} {85} (\bibinfo {year} {2005})}\BibitemShut {NoStop}%
\bibitem [{\citenamefont {Pati}\ \emph {et~al.}(2007)\citenamefont {Pati},
  \citenamefont {Salit}, \citenamefont {Salit},\ and\ \citenamefont
  {Shahriar}}]{Pati07PRL}%
  \BibitemOpen
  \bibfield  {author} {\bibinfo {author} {\bibfnamefont {G.~S.}\ \bibnamefont
  {Pati}}, \bibinfo {author} {\bibfnamefont {M.}~\bibnamefont {Salit}},
  \bibinfo {author} {\bibfnamefont {K.}~\bibnamefont {Salit}},\ and\ \bibinfo
  {author} {\bibfnamefont {M.~S.}\ \bibnamefont {Shahriar}},\ }\bibfield
  {title} {\bibinfo {title} {Demonstration of a tunable-bandwidth white-light
  interferometer using anomalous dispersion in atomic vapor},\ }\href@noop {}
  {\bibfield  {journal} {\bibinfo  {journal} {Phys. Rev. Lett.}\ }\textbf
  {\bibinfo {volume} {99}},\ \bibinfo {pages} {133601} (\bibinfo {year}
  {2007})}\BibitemShut {NoStop}%
\bibitem [{\citenamefont {Wu}\ and\ \citenamefont {Xiao}(2008)}]{Wu08PRA}%
  \BibitemOpen
  \bibfield  {author} {\bibinfo {author} {\bibfnamefont {H.}~\bibnamefont
  {Wu}}\ and\ \bibinfo {author} {\bibfnamefont {M.}~\bibnamefont {Xiao}},\
  }\bibfield  {title} {\bibinfo {title} {White-light cavity with competing
  linear and nonlinear dispersions},\ }\href@noop {} {\bibfield  {journal}
  {\bibinfo  {journal} {Phys. Rev. A}\ }\textbf {\bibinfo {volume} {77}},\
  \bibinfo {pages} {031801(R)} (\bibinfo {year} {2008})}\BibitemShut {NoStop}%
\bibitem [{\citenamefont {Yum}\ \emph {et~al.}(2013{\natexlab{a}})\citenamefont
  {Yum}, \citenamefont {Sheuer}, \citenamefont {Salit}, \citenamefont
  {Hemmer},\ and\ \citenamefont {Shahriar}}]{Yum13JLT}%
  \BibitemOpen
  \bibfield  {author} {\bibinfo {author} {\bibfnamefont {H.~N.}\ \bibnamefont
  {Yum}}, \bibinfo {author} {\bibfnamefont {J.}~\bibnamefont {Sheuer}},
  \bibinfo {author} {\bibfnamefont {M.}~\bibnamefont {Salit}}, \bibinfo
  {author} {\bibfnamefont {P.~R.}\ \bibnamefont {Hemmer}},\ and\ \bibinfo
  {author} {\bibfnamefont {M.~S.}\ \bibnamefont {Shahriar}},\ }\bibfield
  {title} {\bibinfo {title} {Demonstration of white light cavity effect using
  stimulated {B}rillouin scattering in a fiber loop},\ }\href@noop {}
  {\bibfield  {journal} {\bibinfo  {journal} {J. Lightwave Technol.}\ }\textbf
  {\bibinfo {volume} {32}},\ \bibinfo {pages} {3865} (\bibinfo {year}
  {2013}{\natexlab{a}})}\BibitemShut {NoStop}%
\bibitem [{\citenamefont {Wise}\ \emph {et~al.}(2005)\citenamefont {Wise},
  \citenamefont {Quetschke}, \citenamefont {Deshpande}, \citenamefont
  {Mueller}, \citenamefont {Reitze}, \citenamefont {Tanner}, \citenamefont
  {Whiting}, \citenamefont {Chen}, \citenamefont {T{\"u}nnermann},
  \citenamefont {Kley},\ and\ \citenamefont {Clausnitzer}}]{Wise05PRL}%
  \BibitemOpen
  \bibfield  {author} {\bibinfo {author} {\bibfnamefont {S.}~\bibnamefont
  {Wise}}, \bibinfo {author} {\bibfnamefont {V.}~\bibnamefont {Quetschke}},
  \bibinfo {author} {\bibfnamefont {A.~J.}\ \bibnamefont {Deshpande}}, \bibinfo
  {author} {\bibfnamefont {G.}~\bibnamefont {Mueller}}, \bibinfo {author}
  {\bibfnamefont {D.~H.}\ \bibnamefont {Reitze}}, \bibinfo {author}
  {\bibfnamefont {D.~B.}\ \bibnamefont {Tanner}}, \bibinfo {author}
  {\bibfnamefont {B.~F.}\ \bibnamefont {Whiting}}, \bibinfo {author}
  {\bibfnamefont {Y.}~\bibnamefont {Chen}}, \bibinfo {author} {\bibfnamefont
  {A.}~\bibnamefont {T{\"u}nnermann}}, \bibinfo {author} {\bibfnamefont
  {E.}~\bibnamefont {Kley}},\ and\ \bibinfo {author} {\bibfnamefont
  {T.}~\bibnamefont {Clausnitzer}},\ }\bibfield  {title} {\bibinfo {title}
  {Phase effects in the diffraction of light: {B}eyond the grating equation},\
  }\href@noop {} {\bibfield  {journal} {\bibinfo  {journal} {Phys. Rev. Lett.}\
  }\textbf {\bibinfo {volume} {95}},\ \bibinfo {pages} {013901} (\bibinfo
  {year} {2005})}\BibitemShut {NoStop}%
\bibitem [{\citenamefont {Yum}\ \emph {et~al.}(2013{\natexlab{b}})\citenamefont
  {Yum}, \citenamefont {Liu}, \citenamefont {Hemmer}, \citenamefont {Scheuer},\
  and\ \citenamefont {Shahriar}}]{Yum13OC}%
  \BibitemOpen
  \bibfield  {author} {\bibinfo {author} {\bibfnamefont {H.~N.}\ \bibnamefont
  {Yum}}, \bibinfo {author} {\bibfnamefont {X.}~\bibnamefont {Liu}}, \bibinfo
  {author} {\bibfnamefont {P.~R.}\ \bibnamefont {Hemmer}}, \bibinfo {author}
  {\bibfnamefont {J.}~\bibnamefont {Scheuer}},\ and\ \bibinfo {author}
  {\bibfnamefont {M.~S.}\ \bibnamefont {Shahriar}},\ }\bibfield  {title}
  {\bibinfo {title} {The fundamental limitations on the practical realizations
  of white light cavities},\ }\href@noop {} {\bibfield  {journal} {\bibinfo
  {journal} {Opt. Commun.}\ }\textbf {\bibinfo {volume} {305}},\ \bibinfo
  {pages} {260} (\bibinfo {year} {2013}{\natexlab{b}})}\BibitemShut {NoStop}%
\bibitem [{\citenamefont {Xu}\ \emph {et~al.}(2007)\citenamefont {Xu},
  \citenamefont {Dong},\ and\ \citenamefont {Lipson}}]{Xu06NPhys}%
  \BibitemOpen
  \bibfield  {author} {\bibinfo {author} {\bibfnamefont {Q.}~\bibnamefont
  {Xu}}, \bibinfo {author} {\bibfnamefont {P.}~\bibnamefont {Dong}},\ and\
  \bibinfo {author} {\bibfnamefont {M.}~\bibnamefont {Lipson}},\ }\bibfield
  {title} {\bibinfo {title} {Breaking the delay-bandwidth limit in a photonic
  structure},\ }\href@noop {} {\bibfield  {journal} {\bibinfo  {journal} {Nat.
  Phys.}\ }\textbf {\bibinfo {volume} {3}},\ \bibinfo {pages} {406} (\bibinfo
  {year} {2007})}\BibitemShut {NoStop}%
\bibitem [{\citenamefont {Shabahang}\ \emph {et~al.}(2017)\citenamefont
  {Shabahang}, \citenamefont {Kondakci}, \citenamefont {Villinger},
  \citenamefont {Perlstein}, \citenamefont {{El H}alawany},\ and\ \citenamefont
  {Abouraddy}}]{Shabahang17SR}%
  \BibitemOpen
  \bibfield  {author} {\bibinfo {author} {\bibfnamefont {S.}~\bibnamefont
  {Shabahang}}, \bibinfo {author} {\bibfnamefont {H.~E.}\ \bibnamefont
  {Kondakci}}, \bibinfo {author} {\bibfnamefont {M.~L.}\ \bibnamefont
  {Villinger}}, \bibinfo {author} {\bibfnamefont {J.~D.}\ \bibnamefont
  {Perlstein}}, \bibinfo {author} {\bibfnamefont {A.}~\bibnamefont {{El
  H}alawany}},\ and\ \bibinfo {author} {\bibfnamefont {A.~F.}\ \bibnamefont
  {Abouraddy}},\ }\bibfield  {title} {\bibinfo {title} {Omni-resonant optical
  micro-cavity},\ }\href@noop {} {\bibfield  {journal} {\bibinfo  {journal}
  {Sci. Rep.}\ }\textbf {\bibinfo {volume} {7}},\ \bibinfo {pages} {10336}
  (\bibinfo {year} {2017})}\BibitemShut {NoStop}%
\bibitem [{\citenamefont {Villinger}\ \emph {et~al.}(2021)\citenamefont
  {Villinger}, \citenamefont {Shiri}, \citenamefont {Shabahang}, \citenamefont
  {Jahromi}, \citenamefont {Nasr}, \citenamefont {Villinger},\ and\
  \citenamefont {Abouraddy}}]{Villinger21AOM}%
  \BibitemOpen
  \bibfield  {author} {\bibinfo {author} {\bibfnamefont {M.~L.}\ \bibnamefont
  {Villinger}}, \bibinfo {author} {\bibfnamefont {A.}~\bibnamefont {Shiri}},
  \bibinfo {author} {\bibfnamefont {S.}~\bibnamefont {Shabahang}}, \bibinfo
  {author} {\bibfnamefont {A.~K.}\ \bibnamefont {Jahromi}}, \bibinfo {author}
  {\bibfnamefont {M.~B.}\ \bibnamefont {Nasr}}, \bibinfo {author}
  {\bibfnamefont {C.}~\bibnamefont {Villinger}},\ and\ \bibinfo {author}
  {\bibfnamefont {A.~F.}\ \bibnamefont {Abouraddy}},\ }\bibfield  {title}
  {\bibinfo {title} {Doubling the near-infrared photocurrent in a solar cell
  via omni-resonant coherent perfect absorption},\ }\href@noop {} {\bibfield
  {journal} {\bibinfo  {journal} {Adv. Opt. Mat.}\ }\textbf {\bibinfo {volume}
  {9}},\ \bibinfo {pages} {2001107} (\bibinfo {year} {2021})}\BibitemShut
  {NoStop}%
\bibitem [{\citenamefont {Jahromi}\ \emph {et~al.}(2021)\citenamefont
  {Jahromi}, \citenamefont {Villinger}, \citenamefont {{El H}alawany},
  \citenamefont {Shabahang}, \citenamefont {Kondakci}, \citenamefont
  {Perlstein},\ and\ \citenamefont {Abouraddy}}]{Jahromi21arxiv}%
  \BibitemOpen
  \bibfield  {author} {\bibinfo {author} {\bibfnamefont {A.~K.}\ \bibnamefont
  {Jahromi}}, \bibinfo {author} {\bibfnamefont {M.~L.}\ \bibnamefont
  {Villinger}}, \bibinfo {author} {\bibfnamefont {A.}~\bibnamefont {{El
  H}alawany}}, \bibinfo {author} {\bibfnamefont {S.}~\bibnamefont {Shabahang}},
  \bibinfo {author} {\bibfnamefont {H.~E.}\ \bibnamefont {Kondakci}}, \bibinfo
  {author} {\bibfnamefont {J.~D.}\ \bibnamefont {Perlstein}},\ and\ \bibinfo
  {author} {\bibfnamefont {A.~F.}\ \bibnamefont {Abouraddy}},\ }\bibfield
  {title} {\bibinfo {title} {Broadband omni-resonant coherent perfect
  absorption in graphene},\ }\href@noop {} {\bibfield  {journal} {\bibinfo
  {journal} {arXiv:2104.08706}\ } (\bibinfo {year} {2021})}\BibitemShut
  {NoStop}%
\bibitem [{\citenamefont {Shin}\ \emph {et~al.}(2004)\citenamefont {Shin},
  \citenamefont {Yanik}, \citenamefont {Fan}, \citenamefont {Zia},\ and\
  \citenamefont {Brongersma}}]{Shin04APL}%
  \BibitemOpen
  \bibfield  {author} {\bibinfo {author} {\bibfnamefont {H.}~\bibnamefont
  {Shin}}, \bibinfo {author} {\bibfnamefont {M.~F.}\ \bibnamefont {Yanik}},
  \bibinfo {author} {\bibfnamefont {S.}~\bibnamefont {Fan}}, \bibinfo {author}
  {\bibfnamefont {R.}~\bibnamefont {Zia}},\ and\ \bibinfo {author}
  {\bibfnamefont {M.~L.}\ \bibnamefont {Brongersma}},\ }\bibfield  {title}
  {\bibinfo {title} {Omnidirectional resonance in a metal--dielectric--metal
  geometry},\ }\href@noop {} {\bibfield  {journal} {\bibinfo  {journal} {Appl.
  Phys. Lett.}\ }\textbf {\bibinfo {volume} {84}},\ \bibinfo {pages} {4421}
  (\bibinfo {year} {2004})}\BibitemShut {NoStop}%
\bibitem [{\citenamefont {Liu}\ and\ \citenamefont
  {Brongersma}(2007)}]{Liu07APL}%
  \BibitemOpen
  \bibfield  {author} {\bibinfo {author} {\bibfnamefont {J.~S.~Q.}\
  \bibnamefont {Liu}}\ and\ \bibinfo {author} {\bibfnamefont {M.~L.}\
  \bibnamefont {Brongersma}},\ }\bibfield  {title} {\bibinfo {title}
  {Omnidirectional light emission via surface plasmon polariton},\ }\href@noop
  {} {\bibfield  {journal} {\bibinfo  {journal} {Appl. Phys. Lett.}\ }\textbf
  {\bibinfo {volume} {90}},\ \bibinfo {pages} {09116} (\bibinfo {year}
  {2007})}\BibitemShut {NoStop}%
\bibitem [{\citenamefont {Shabahang}\ \emph {et~al.}(2019)\citenamefont
  {Shabahang}, \citenamefont {Jahromi}, \citenamefont {Shiri}, \citenamefont
  {Schepler},\ and\ \citenamefont {Abouraddy}}]{Shabahang19OL}%
  \BibitemOpen
  \bibfield  {author} {\bibinfo {author} {\bibfnamefont {S.}~\bibnamefont
  {Shabahang}}, \bibinfo {author} {\bibfnamefont {A.~K.}\ \bibnamefont
  {Jahromi}}, \bibinfo {author} {\bibfnamefont {A.}~\bibnamefont {Shiri}},
  \bibinfo {author} {\bibfnamefont {K.~L.}\ \bibnamefont {Schepler}},\ and\
  \bibinfo {author} {\bibfnamefont {A.~F.}\ \bibnamefont {Abouraddy}},\
  }\bibfield  {title} {\bibinfo {title} {Toggling between active and passive
  imaging with an omni-resonant micro-cavity},\ }\href@noop {} {\bibfield
  {journal} {\bibinfo  {journal} {Opt. Lett.}\ }\textbf {\bibinfo {volume}
  {44}},\ \bibinfo {pages} {1532} (\bibinfo {year} {2019})}\BibitemShut
  {NoStop}%
\bibitem [{\citenamefont {Kondakci}\ and\ \citenamefont
  {Abouraddy}(2017)}]{Kondakci17NP}%
  \BibitemOpen
  \bibfield  {author} {\bibinfo {author} {\bibfnamefont {H.~E.}\ \bibnamefont
  {Kondakci}}\ and\ \bibinfo {author} {\bibfnamefont {A.~F.}\ \bibnamefont
  {Abouraddy}},\ }\bibfield  {title} {\bibinfo {title} {Diffraction-free
  space-time beams},\ }\href@noop {} {\bibfield  {journal} {\bibinfo  {journal}
  {Nat. Photon.}\ }\textbf {\bibinfo {volume} {11}},\ \bibinfo {pages} {733}
  (\bibinfo {year} {2017})}\BibitemShut {NoStop}%
\bibitem [{\citenamefont {Kondakci}\ and\ \citenamefont
  {Abouraddy}(2019)}]{Kondakci19NC}%
  \BibitemOpen
  \bibfield  {author} {\bibinfo {author} {\bibfnamefont {H.~E.}\ \bibnamefont
  {Kondakci}}\ and\ \bibinfo {author} {\bibfnamefont {A.~F.}\ \bibnamefont
  {Abouraddy}},\ }\bibfield  {title} {\bibinfo {title} {Optical space-time wave
  packets of arbitrary group velocity in free space},\ }\href@noop {}
  {\bibfield  {journal} {\bibinfo  {journal} {Nat. Commun.}\ }\textbf {\bibinfo
  {volume} {10}},\ \bibinfo {pages} {929} (\bibinfo {year} {2019})}\BibitemShut
  {NoStop}%
\bibitem [{\citenamefont {Yessenov}\ \emph {et~al.}(2022)\citenamefont
  {Yessenov}, \citenamefont {Hall}, \citenamefont {Schepler},\ and\
  \citenamefont {Abouraddy}}]{Yessenov22AOP}%
  \BibitemOpen
  \bibfield  {author} {\bibinfo {author} {\bibfnamefont {M.}~\bibnamefont
  {Yessenov}}, \bibinfo {author} {\bibfnamefont {L.~A.}\ \bibnamefont {Hall}},
  \bibinfo {author} {\bibfnamefont {K.~L.}\ \bibnamefont {Schepler}},\ and\
  \bibinfo {author} {\bibfnamefont {A.~F.}\ \bibnamefont {Abouraddy}},\
  }\bibfield  {title} {\bibinfo {title} {Space-time wave packets},\ }\href@noop
  {} {\bibfield  {journal} {\bibinfo  {journal} {arXiv:2201.08297}\ } (\bibinfo
  {year} {2022})}\BibitemShut {NoStop}%
\bibitem [{\citenamefont {Shiri}\ \emph
  {et~al.}(2020{\natexlab{a}})\citenamefont {Shiri}, \citenamefont {Yessenov},
  \citenamefont {Aravindakshan},\ and\ \citenamefont {Abouraddy}}]{Shiri20OL}%
  \BibitemOpen
  \bibfield  {author} {\bibinfo {author} {\bibfnamefont {A.}~\bibnamefont
  {Shiri}}, \bibinfo {author} {\bibfnamefont {M.}~\bibnamefont {Yessenov}},
  \bibinfo {author} {\bibfnamefont {R.}~\bibnamefont {Aravindakshan}},\ and\
  \bibinfo {author} {\bibfnamefont {A.~F.}\ \bibnamefont {Abouraddy}},\
  }\bibfield  {title} {\bibinfo {title} {Omni-resonant space-time wave
  packets},\ }\href@noop {} {\bibfield  {journal} {\bibinfo  {journal} {Opt.
  Lett.}\ }\textbf {\bibinfo {volume} {45}},\ \bibinfo {pages} {1774} (\bibinfo
  {year} {2020}{\natexlab{a}})}\BibitemShut {NoStop}%
\bibitem [{\citenamefont {Shiri}\ \emph
  {et~al.}(2020{\natexlab{b}})\citenamefont {Shiri}, \citenamefont {Schepler},\
  and\ \citenamefont {Abouraddy}}]{Shiri20APLP}%
  \BibitemOpen
  \bibfield  {author} {\bibinfo {author} {\bibfnamefont {A.}~\bibnamefont
  {Shiri}}, \bibinfo {author} {\bibfnamefont {K.~L.}\ \bibnamefont
  {Schepler}},\ and\ \bibinfo {author} {\bibfnamefont {A.~F.}\ \bibnamefont
  {Abouraddy}},\ }\bibfield  {title} {\bibinfo {title} {Programmable
  omni-resonance using space-time fields},\ }\href@noop {} {\bibfield
  {journal} {\bibinfo  {journal} {APL Photon.}\ }\textbf {\bibinfo {volume}
  {5}},\ \bibinfo {pages} {106107} (\bibinfo {year}
  {2020}{\natexlab{b}})}\BibitemShut {NoStop}%
\bibitem [{\citenamefont {Saleh}\ and\ \citenamefont
  {Teich}(2007)}]{SalehBook07}%
  \BibitemOpen
  \bibfield  {author} {\bibinfo {author} {\bibfnamefont {B.~E.~A.}\
  \bibnamefont {Saleh}}\ and\ \bibinfo {author} {\bibfnamefont {M.~C.}\
  \bibnamefont {Teich}},\ }\href@noop {} {\emph {\bibinfo {title} {Principles
  of Photonics}}}\ (\bibinfo  {publisher} {Wiley},\ \bibinfo {year}
  {2007})\BibitemShut {NoStop}%
\bibitem [{\citenamefont {England}\ \emph {et~al.}(2014)\citenamefont
  {England}, \citenamefont {Kolle}, \citenamefont {Kim}, \citenamefont {Khan},
  \citenamefont {Mu{\~n}oz}, \citenamefont {Mazur},\ and\ \citenamefont
  {Aizenberg}}]{England14PNAS}%
  \BibitemOpen
  \bibfield  {author} {\bibinfo {author} {\bibfnamefont {G.}~\bibnamefont
  {England}}, \bibinfo {author} {\bibfnamefont {M.}~\bibnamefont {Kolle}},
  \bibinfo {author} {\bibfnamefont {P.}~\bibnamefont {Kim}}, \bibinfo {author}
  {\bibfnamefont {M.}~\bibnamefont {Khan}}, \bibinfo {author} {\bibfnamefont
  {P.}~\bibnamefont {Mu{\~n}oz}}, \bibinfo {author} {\bibfnamefont
  {E.}~\bibnamefont {Mazur}},\ and\ \bibinfo {author} {\bibfnamefont
  {J.}~\bibnamefont {Aizenberg}},\ }\bibfield  {title} {\bibinfo {title}
  {Bioinspired micrograting arrays mimicking the reverse color diffraction
  elements evolved by the butterfly \textit{Pierella luna}},\ }\href@noop {}
  {\bibfield  {journal} {\bibinfo  {journal} {Proc. Natl. Acad. Sci. USA}\
  }\textbf {\bibinfo {volume} {111}},\ \bibinfo {pages} {15630} (\bibinfo
  {year} {2014})}\BibitemShut {NoStop}%
\bibitem [{\citenamefont {Lin}\ \emph {et~al.}(2015)\citenamefont {Lin},
  \citenamefont {Sun}, \citenamefont {Liu},\ and\ \citenamefont
  {Hu}}]{Lin15OL}%
  \BibitemOpen
  \bibfield  {author} {\bibinfo {author} {\bibfnamefont {H.}~\bibnamefont
  {Lin}}, \bibinfo {author} {\bibfnamefont {X.}~\bibnamefont {Sun}}, \bibinfo
  {author} {\bibfnamefont {J.}~\bibnamefont {Liu}},\ and\ \bibinfo {author}
  {\bibfnamefont {J.}~\bibnamefont {Hu}},\ }\bibfield  {title} {\bibinfo
  {title} {Diffractive broadband coupling into high-{Q} resonant cavities},\
  }\href@noop {} {\bibfield  {journal} {\bibinfo  {journal} {Opt. Lett.}\
  }\textbf {\bibinfo {volume} {40}},\ \bibinfo {pages} {2377} (\bibinfo {year}
  {2015})}\BibitemShut {NoStop}%
\bibitem [{\citenamefont {Arbabi}\ \emph {et~al.}(2017)\citenamefont {Arbabi},
  \citenamefont {Arbabi}, \citenamefont {Kamali}, \citenamefont {Horie},\ and\
  \citenamefont {Faraon}}]{Arbabi17Optica}%
  \BibitemOpen
  \bibfield  {author} {\bibinfo {author} {\bibfnamefont {E.}~\bibnamefont
  {Arbabi}}, \bibinfo {author} {\bibfnamefont {A.}~\bibnamefont {Arbabi}},
  \bibinfo {author} {\bibfnamefont {S.~M.}\ \bibnamefont {Kamali}}, \bibinfo
  {author} {\bibfnamefont {Y.}~\bibnamefont {Horie}},\ and\ \bibinfo {author}
  {\bibfnamefont {A.}~\bibnamefont {Faraon}},\ }\bibfield  {title} {\bibinfo
  {title} {Controlling the sign of chromatic dispersion in diffractive optics
  with dielectric metasurfaces},\ }\href@noop {} {\bibfield  {journal}
  {\bibinfo  {journal} {Optica}\ }\textbf {\bibinfo {volume} {4}},\ \bibinfo
  {pages} {625} (\bibinfo {year} {2017})}\BibitemShut {NoStop}%
\bibitem [{\citenamefont {{McClung}}\ \emph {et~al.}(2020)\citenamefont
  {{McClung}}, \citenamefont {Mansouree},\ and\ \citenamefont
  {Arbabi}}]{McClung20Light}%
  \BibitemOpen
  \bibfield  {author} {\bibinfo {author} {\bibfnamefont {A.}~\bibnamefont
  {{McClung}}}, \bibinfo {author} {\bibfnamefont {M.}~\bibnamefont
  {Mansouree}},\ and\ \bibinfo {author} {\bibfnamefont {A.}~\bibnamefont
  {Arbabi}},\ }\bibfield  {title} {\bibinfo {title} {At-will chromatic
  dispersion by prescribing light trajectories with cascaded metasurfaces},\
  }\href@noop {} {\bibfield  {journal} {\bibinfo  {journal} {Light: Sci.
  Appl.}\ }\textbf {\bibinfo {volume} {9}},\ \bibinfo {pages} {93} (\bibinfo
  {year} {2020})}\BibitemShut {NoStop}%
\end{thebibliography}%

%\bibliographyfullrefs{diffraction}

\end{document}